\documentclass[twocolumns]{aa}
\usepackage{graphicx}
\usepackage{txfonts}
\begin{document}
   \title{The spatial clustering of X-ray selected AGN and galaxies in
the Chandra Deep Field South and North }

\author{
R. Gilli\inst{1}, 
E. Daddi\inst{2},
G. Zamorani\inst{3},
P. Tozzi\inst{4},
S. Borgani\inst{5}, 
J. Bergeron\inst{6}, 
R. Giacconi\inst{7,8},
G. Hasinger\inst{9}, 
V. Mainieri\inst{9},
C. Norman\inst{7,10}, 
P. Rosati\inst{2},
G. Szokoly\inst{9}, 
W. Zheng\inst{10}
}

\authorrunning{R. Gilli et al.}
\titlerunning{Spatial clustering of X-ray selected sources}

   \offprints{R. Gilli \\email:{\tt gilli@arcetri.astro.it}}

   \date{Received ... ; accepted ...}

\institute{
Istituto Nazionale di Astrofisica (INAF) - Osservatorio Astrofisico di
Arcetri, Largo E. Fermi 5, 50125 Firenze, Italy
\and
European Southern Observatory, Karl-Schwarzschild-Strasse 2, Garching,
D-85748, Germany
\and 
Istituto Nazionale di Astrofisica (INAF) - Osservatorio Astronomico di
Bologna, Via Ranzani 1, 40127 Bologna, Italy
\and 
Istituto Nazionale di Astrofisica (INAF) - Osservatorio Astronomico,
Via G. Tiepolo 11, 34131 Trieste, Italy
\and 
INFN, c/o Dip. di Astronomia dell'Universit\`a, Via G. Tiepolo 11,
34131 Trieste, Italy
\and 
Institut d'Astrophysique de Paris, 98bis Boulevard Arago, 75014 Paris,
France
\and 
The Johns Hopkins University, Homewood Campus, Baltimore, MD 21218
\and 
Associated Universities, Inc., 1400 16th Street NW, Suite 730,
Washington, DC 20036
\and
Max-Planck-Institut f\"ur extraterrestrische Physik, Postfach 1312,
D-85741 Garching, Germany
\and 
Space Telescope Science Institute, 3700 San Martin Drive, Baltimore,
MD 21218
}

\abstract{ We investigate the spatial clustering of X-ray selected
sources in the two deepest X-ray fields to date, namely the 2Msec {\it
Chandra} Deep Field North (CDFN) and the 1Msec {\it Chandra} Deep
Field South (CDFS). The projected correlation function $w(r_p)$,
measured on scales $\sim 0.2-10 \:h^{-1}$ Mpc for a sample of 240
sources with spectroscopic redshift in the CDFN and 124 sources in the
CDFS at a median redshift of $\bar z\sim0.8$, is used to constrain the
amplitude and slope of the real space correlation function
$\xi(r)=(r/r_0)^{-\gamma}$. The clustering signal is detected at
high confidence ($\gtrsim 7 \sigma$) in both fields. The amplitude of the
correlation is found to be significantly different in the two fields,
the correlation length $r_0$ being $8.6 \pm 1.2\:h^{-1}$ Mpc in the
CDFS and $4.2\pm 0.4\:h^{-1}$ Mpc in the CDFN, while the correlation
slope $\gamma$ is found to be flat in both fields: $\gamma=1.33\pm
0.11 $ in the CDFS and $\gamma=1.42\pm 0.07 $ in the CDFN (a flat
Universe with $\Omega_m=0.3$ and $\Omega_{\Lambda}=0.7$ is assumed;
$1\sigma$ Poisson error estimates are considered).
The correlation function has been also measured separately for sources
classified as AGN or galaxies. In both fields AGN have a median
redshift of $\bar z\sim0.9$ and a median 0.5-10 keV luminosity of
$\bar L_x\sim 10^{43}$ erg~s$^{-1}$, i.e. they are generally in the Seyfert
luminosity regime.
As in the case of the total samples, we found a significant difference
in the AGN clustering amplitude between the two fields, the best fit
correlation parameters being $r_0=10.3\pm 1.7 \:h^{-1}$ Mpc,
$\gamma=1.33 \pm 0.14 $ in the CDFS, and $r_0=5.5\pm 0.6 \:h^{-1}$ Mpc,
$\gamma=1.50 \pm 0.12 $ in the CDFN. In the
CDFN, where the statistics is sufficiently high, we were also able to
measure the clustering of X-ray selected galaxies, finding $r_0=4.0\pm
0.7 \:h^{-1}$ Mpc and $\gamma=1.36 \pm 0.15 $.
Within each field no statistically significant difference is found
between soft and hard X-ray selected sources or between type 1 and
type 2 AGN. 
After having discussed and ruled out the possibility that the observed
variance in the clustering amplitude be due to observational biases,
we verified that the extra correlation signal in the CDFS is primarily
due to the two prominent redshift spikes at $z\sim 0.7$ reported by
Gilli et al. (\cite{gilli03}).
The high ($5-10\:h^{-1}$ Mpc) correlation length measured for the
X-ray selected AGN at $z\sim 1$ in the two {\it Chandra} Msec fields
is comparable to that of early type galaxies at the same redshift.
This is consistent with the idea that, at $z\sim 1$, AGN with
Seyfert-like luminosities might be generally hosted by massive
galaxies.

   }

   \maketitle

\section{Introduction} \label{introduction}

Active Galactic Nuclei (AGN) represent one of the best tools to study
the large scale structure of the Universe at intermediate-high
redshifts, $z\sim 1-2$, i.e. at an epoch of intense structure
formation where matter was undergoing the transition from the
initially smooth state observed at the recombination ($z\sim1000$) to
the clumpy distribution observed at present time (see e.g. Hartwick \&
Schade 1990).

One of the most commonly used statistics to measure the clustering of
a population of sources is the two-point correlation function
$\xi(r)$, which measures the excess probability of finding a pair of
objects at a separation $r$ with respect to a random distribution and
is usually approximated by a power law
$\xi(r)=(r/r_0)^{-\gamma}$. Under simple assumptions, the amplitude of
the AGN correlation function can be used to estimate the typical mass
of the dark matter halos in which AGN reside (Grazian et
al. \cite{grazi04}, Magliocchetti et al. \cite{maglio04}) and the
typical AGN lifetimes (Martini \& Weinberg \cite{marti01}).


The first attempts to measure AGN clustering date more than 20 years
ago (Osmer \cite{osmer81}). Since then AGN clustering has been
extensively studied and detected by means of optical surveys
encompassing an increasing number of QSOs (Shanks et
al. \cite{shank87}, La Franca et al. \cite{lafra98}, Croom et
al. \cite{croom01}, Grazian et al. \cite{grazi04}). Recently, the 2dF
QSO Redshift Survey (2QZ, Croom et al. \cite{croom01}) has provided
the tightest constraints to QSO clustering, based on a sample of more
than $10^4$ objects: the QSO correlation length and slope were found
to be $r_0=5.7 \pm 0.5 \:h^{-1}$ Mpc and $\gamma=1.56\pm 0.10$ at a
median redshift of $\bar z=1.5$ and on comoving scales of $1-60 \:
h^{-1}$ Mpc. This result confirmed previous measurements and showed
that QSO clustering at $z=1.5$ is comparable to that of local ($z \sim
0.05$) optically selected galaxies (Tucker et al. \cite{tuck97},
Ratcliffe et al. \cite{rat98}). In addition, thanks to the large
number of QSOs in their sample, Croom et al. (\cite{croom01}) were also
able to investigate the evolution of QSO clustering with redshift,
finding a marginal increase by a factor of 1.4 in the $r_0$ value from
$z\sim0.7$ to $z\sim2.4$ for a flat cosmology with $\Omega_m=0.3$ and
$\Omega_\Lambda=0.7$.
                    
Although optical surveys provide the largest AGN samples so far, they
include almost exclusively unobscured-type 1 objects, since AGN
candidates are mainly selected by means of UV excess techniques.
Obscured-type 2 AGN might instead be efficiently selected by means of
mid- and far-infrared surveys, since the nuclear UV radiation absorbed
by the obscuring medium is expected to be re-emitted at longer
wavelengths. Georgantopoulos \& Shanks (\cite{geo94}) analyzed the
clustering properties of a sample of $\sim 200$ local Seyfert galaxies
($z<0.1$) observed with IRAS and selected through their warm infrared
colors. By comparing the observed number of independent pairs with
that expected from a random sample distributed over the same scales,
they measured a $\sim 3\sigma$ clustering signal for the total sample,
finding marginal evidence that Seyfert 2 galaxies are more clustered
than Seyfert 1s.


Perhaps the most efficient way to sample the obscured AGN population
is through X-ray observations, especially in the hard band, where the
nuclear radiation is less affected by absorption. Based on population
synthesis models for the X-ray background (e.g. Comastri et
al. \cite{comas95}, Gilli et al. \cite{gilli01}, Ueda et
al. \cite{ueda03}), obscured AGN are believed to be a factor of
$\gtrsim 4$ more abundant than unobscured ones and should therefore
dominate the whole AGN population.
Spatial clustering of X-ray selected AGN has been limited so far by
the lack of sizable samples of optically identified X-ray
sources. Boyle \& Mo (\cite{boyle93}) studied the AGN at $z<0.2$ in
the Einstein Medium Sensitivity Survey (EMSS, Stocke et al. 1991),
without finding any positive clustering signal. Carrera et
al. (\cite{carrera98}) considered the AGN in the ROSAT International
X-ray Optical Survey (RIXOS, Mason et al. 2000) and in the Deep ROSAT
Survey (DRS, Boyle et al. \cite{boyle94}), detecting only a weak
($\sim 2\sigma$) clustering signal on scales $<40-80\:h^{-1}$ Mpc for
the RIXOS AGN subsample in the redshift range $z=0.5-1.0$.
Significant clustering signal was instead detected from angular
correlations by several Authors: Akylas et al. (2000), based on the
ROSAT All Sky Survey (RASS, Voges et al. 1999); Vikhlinin \& Forman
(\cite{vikh95}) from a compilation of ROSAT PSPC deep pointings, and
finally Giacconi et al. (2001) from the first 130 ksec observation of
the {\it Chandra} Deep Field South (Rosati et al. 2002). Very recently
Yang et al. (\cite{yang03}) have claimed that hard X-ray selected
sources have an angular clustering amplitude ten times higher than
that of soft X-ray selected sources. A high angular clustering
amplitude for hard X-ray selected sources, consistent with that
measured by Yang et al. (\cite{yang03}), has been also measured by
Basilakos et al. (\cite{basil04}). In some cases (e.g. Vikhlinin \&
Forman 1995; Akylas et al. 2000; Basilakos et al. 2004) the angular
clustering was converted to spatial clustering by means of the
Limber's equation, where an {\it a priori} redshift distribution
has to be assumed. Unfortunately, because of the several uncertainties
in its assumptions, this method has not provided stringent results:
Akylas et al. (\cite{akylas00}) found $r_0=5-8\: h^{-1}$ Mpc,
Vikhlinin \& Forman (\cite{vikh95}) $r_0\gtrsim 5\: h^{-1}$ Mpc and
Basilakos et al.  (\cite{basil04}) $r_0\gtrsim 9\: h^{-1}$ Mpc.

To date, the only direct measurement of spatial clustering of X-ray
selected AGN has been obtained from the ROSAT North Ecliptic Pole
survey data (NEP, Gioia et al. \cite{gioia03}). From a sample of
219 soft X-ray selected AGN, Mullis et al. (\cite{mullis04}) measured a
correlation length of $r_0=7.4^{+1.8}_{-1.9}\:h^{-1}$ Mpc with
$\gamma$ fixed to 1.8. The median redshift of the NEP AGN contributing
to the clustering signal is $\bar z\sim0.2$ (see also Mullis
\cite{mullis01} for a preliminary version of that work). Because of
the relatively short exposures in the NEP survey and the limited ROSAT
sensitivity, only bright sources, with a surface density of the order
of 3 deg$^{-2}$, were detected in this sample. In deeper samples,
where the source surface density is higher, the clustering signal
should be detected more easily since the spatial correlation function
is a power law increasing at lower pair separations. In particular,
deep pencil beam surveys are expected to provide the highest signal
significance with the minimum number of identified objects.

The {\it Chandra} Msec surveys in the Deep Field South (CDFS, Rosati
et al.  2002) and North (CDFN, Alexander et al. 2003) are in this
respect the ideal fields to look at, with an X-ray source surface
density of the order of $3000-4000$ deg$^{-2}$. The drawbacks are that
these strong signals expected on small areas may be subject to
substantial variance, well beyond the one implied by Poisson
statistics (see Daddi et al. 2001 for a discussion of this effect in
the case of angular clustering), so that the ``real'' amplitude of the
correlation function would need a large set of measurements in
independent fields to be reliably estimated. In addition, optical
spectroscopy is challenging for a significant fraction of these X-ray
sources with faint optical magnitude counterparts. We will address
these points in the rest of the paper.  A large spectroscopic
identification program down to faint magnitudes ($R<25.5$) is underway
in the CDFS (Szokoly et al. \cite{szoko04}) and in the CDFN (Barger et
al. \cite{barger03}). To date, about 40-50\% of the X-ray samples have
been spectroscopically identified, revealing that, even at very low
fluxes, AGN are still the most numerous sources populating the X-ray
sky. Here we will take advantage of the spectroscopically identified
sources in the CDFS and CDFN to measure and compare the spatial
clustering of X-ray selected AGN in the two fields.

The paper is organized as follows. In Section 2 we summarize the X-ray
and optical observations of the CDFS and CDFN and present the source
catalogs used in our analysis. In Section 3 we describe the
classification scheme adopted to divide sources into AGN or
galaxies. In Section 4 we describe the methods used to estimate the
projected correlation function of X-ray selected sources as well as
the obtained results, which are then discussed in Section
5. Conclusions and prospects for future work are finally presented in
Section 6.


Throughout this paper we will use a flat cosmology with $\Omega_m=0.3$
and $\Omega_{\Lambda}=0.7$. Unless otherwise stated, we will always
refer to comoving distances in units of $h^{-1}$ Mpc, where
$H_0=100\;h$ km s$^{-1}$ Mpc$^{-1}$. Luminosities are calculated using
$h=0.7$.

\begin{figure}[t]
\includegraphics[width=9cm]{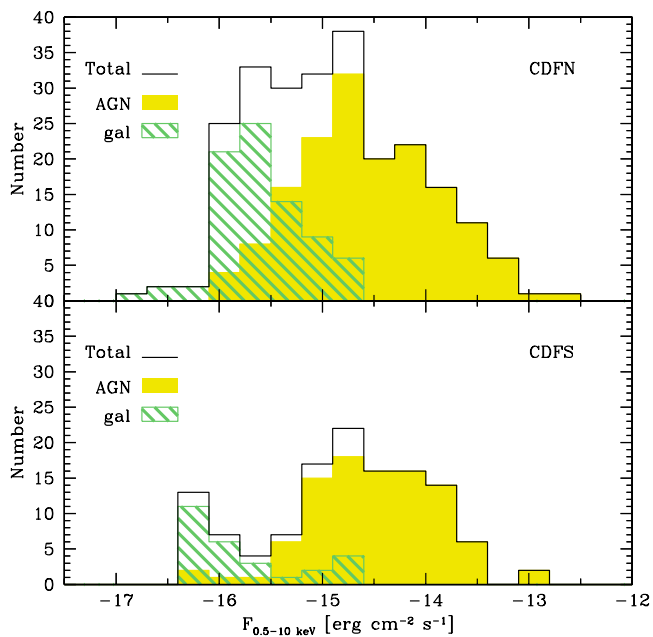}
\caption{X-ray flux distribution for the total, AGN and galaxy sample
observed in the 2Msec CDFN (upper panel) and 1Msec CDFS (lower
panel). Only sources with robust spectroscopic redshift are
considered. The source classification is based on the hardness ratio
vs luminosity diagram described in Section~3 and shown in
Fig.~\ref{cls} and \ref{cln}.}
\label{ftdist}
\end{figure}

\begin{figure}[t]
\includegraphics[width=9cm]{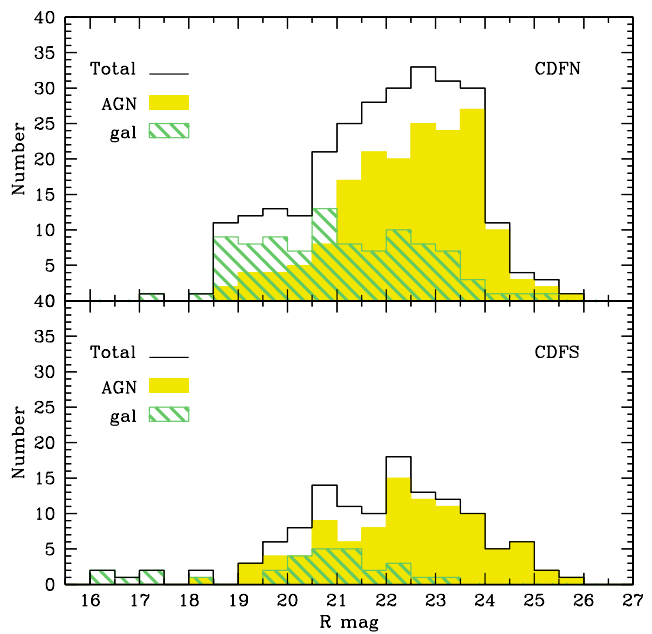}
\caption{R magnitude distribution for the total, AGN and galaxy sample
observed in the 2Msec CDFN (upper panel) and 1Msec CDFS (lower
panel). Only sources with robust spectroscopic redshift are
considered. The source classification is based on the hardness ratio
vs luminosity diagram described in Section~3 and shown in
Fig.~\ref{cls} and \ref{cln}.}
\label{rdist}
\end{figure}

\section{X-ray and optical data} \label{xray}

\subsection {CDFS}

\begin{figure}
\includegraphics[width=9cm]{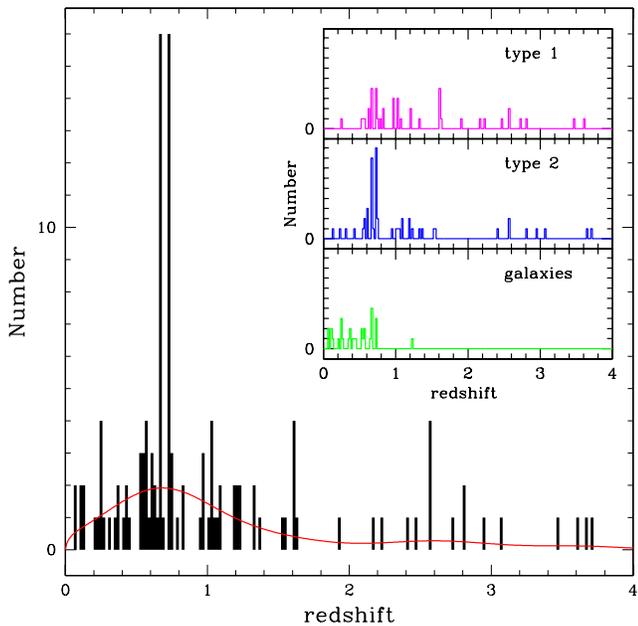}
\caption{Redshift distribution for point-like X-ray sources in the
CDFS in bins of $\Delta z = 0.02$. Only sources with robust
spectroscopic redshift have been considered. The solid curve shows the
selection function obtained by smoothing the observed redshift
distribution. The inset shows the redshift distribution of CDFS
sources as a function of their classification (see Section~3).}
\label{zcdfs}
\end{figure}

The CDFS has been observed with 11 ACIS-I pointings for a total 1 Msec
exposure (Rosati et al. \cite{rosati02}). X-ray sources have been
detected down to limiting fluxes of $5.5\;10^{-17}$ erg cm$^{-2}$
s$^{-1}$ (hereafter cgs) and $4.5\;10^{-16}$ cgs in the soft (0.5-2
keV) and hard (2-10 keV) band, respectively. Overall, 307 sources have
been detected in the soft band and 251 sources in the hard band for a
total sample of 346 sources distributed over the whole 0.1 deg$^2$
field.  The full X-ray catalog and the details of the detection
process have been presented by Giacconi et al. (\cite{giac02}). The
optical follow-up photometry was primarily performed using the FORS1
camera at the VLT (Szokoly et al. \cite{szoko04}). The combined R band
data cover a $13.6 \times 13.6$ arcmin field to limiting magnitudes
between 26 and 26.7. In the area not covered by FORS mosaics, we used
shallower data from the ESO Imaging Survey (EIS, Arnouts et
al. 2001). The optical identification process is described in Tozzi et
al. (\cite{tozzi01}) and Giacconi et al. (\cite{giac02}). Optical
spectroscopy for most of the X-ray counterparts with R$<24$ has been
obtained with FORS1 during several observational runs at the
VLT. About $\sim 20$ spectra of optically faint sources with $24<R<26$
were also collected.  The details of the spectroscopic data reduction
and analysis are presented in Szokoly et al.  (\cite{szoko04}). So far
169 redshifts have been obtained. Quality flags have been assigned to
the spectra, according to their reliability. Here we consider only the
127 X-ray point-like sources (excluding stars) with spectral quality
flag $Q\geq2$, where two or more lines have been observed in the
spectrum of the optical counterpart and the redshift determination is
unambiguous. The X-ray flux and R band magnitude distribution for
these sources are shown in the lower panels of Fig.~\ref{ftdist} and
Fig.~\ref{rdist}, respectively. We estimated the redshift accuracy by
considering the $\sim 40$ sources with at least two independent
redshift measurements, both with $Q\geq2$, obtained in different
observing runs (see Table 5 of Szokoly et al. 2004). The distribution
of the redshift differences has a relatively large dispersion of
$\sigma(\Delta\:z)\sim0.005$. When removing two outliers with a
$3\sigma$ clipping technique (both outliers are Broad Lines AGN for
which a precise redshift determination is more difficult), the
observed dispersion decreases to $\sigma(\Delta\:z)\sim0.003$,
corresponding to an average uncertainty in a single redshift
measurement of $\Delta\:z\sim0.003/\sqrt 2\sim0.002$.\footnote{We note
that the value of 0.005 quoted by Szokoly et al. (\cite{szoko04}) as
the typical uncertainty in the redshift determination is a
conservative $\sim3\sigma$ boundary.} As shown in Fig.~\ref{zcdfs} the
redshift distribution is dominated by two large concentrations of
sources at z=0.67 and z=0.73, while other smaller peaks are also
visible (see also Gilli et al. 2003), already demonstrating that X-ray
sources in the CDFS are highly clustered. The final spectroscopic
completeness is $\sim 35\%$. This fraction increases to 78\% for the
subsample of X-ray sources with optical counterparts brighter than
R=24. We stress that in our measurements it is essential to consider
only sources with small redshift errors, otherwise the clustering
signal in redshift space would be removed. The typical measurement
errors in the photometric redshifts of CDFS sources (Zheng et
al. \cite{zheng04}) are of the order of $\Delta z\sim0.14$,
corresponding to $\sim 270 h^{-1}$ Mpc comoving at the median CDFS
redshift of 0.7. The above redshift uncertainty would significantly
dilute the clustering signal in the considered field (which is
dominated by redshift clustering) and therefore photometric redshifts
cannot be used for our purposes.

\begin{figure}
\includegraphics[width=9cm]{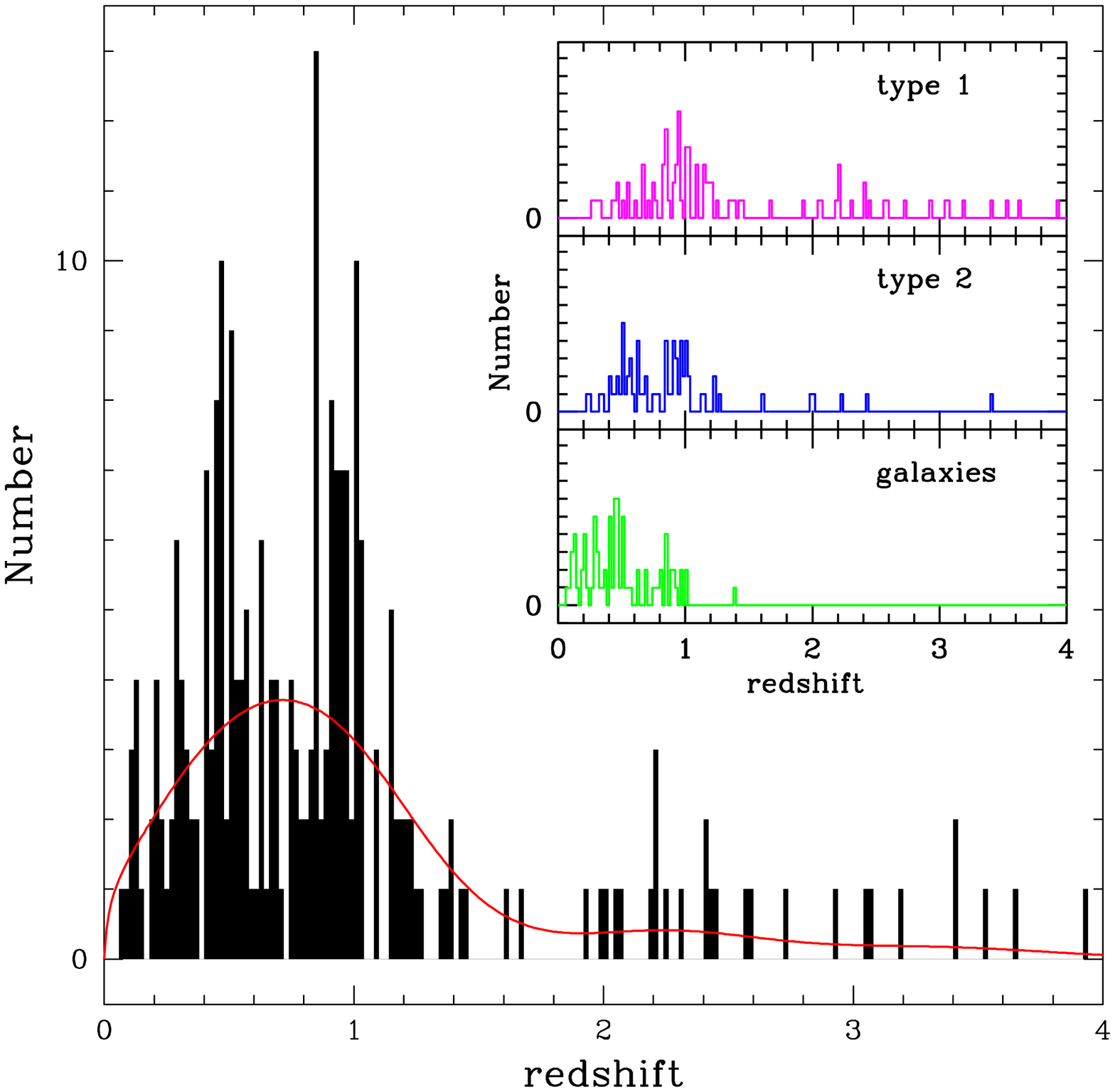}
\caption{Same as Fig.~\ref{zcdfs} but for CDFN sources.}
\label{zcdfn}
\end{figure}

\subsection {CDFN}

The {\it Chandra} Deep Field North (CDFN, Alexander et
al. \cite{alex03}, Barger et al. \cite{barger03}), which is centered
on the Hubble Deep Field North (Williams et al. \cite{will96}), is the
analog of the CDFS in the Northern hemisphere. The CDFN has been
observed with 20 ACIS-I pointings for a total 2 Msec
exposure. Limiting fluxes of $\sim 2.5\;10^{-17}$ cgs and $\sim1.4
\;10^{-16}$ cgs have been reached in the soft and hard band,
respectively. A total sample of 503 X-ray sources (451 of which are
detected in the soft band and 332 in the hard band) has been collected
over an area of 0.13 deg$^2$. The full X-ray catalog is found in
Alexander et al. (\cite{alex03}) and the details of the optical
identification program have been published by Barger et
al. (\cite{barger03}). The LRIS and DEIMOS instruments at the Keck
telescope were primarily used for the optical follow-up of the X-ray
sources. A few additional identifications were added by cross
correlating the X-ray with the optical catalog of the Caltech Faint
Galaxy Redshift Survey (Cohen et al. \cite{cohen00}) which covers the
inner 50 arcmin$^2$ of the CDFN and has a spectroscopic completeness
of about 90\% down to R=24 in the Hubble Deep Field and to R=23 in the
surrounding flanking fields. Most of the redshifts in the Barger et
al. (\cite{barger03}) catalog have been obtained from spectra with
multiple lines, and should be therefore comparable to the $Q\geq2$
redshifts of the CDFS catalog. We ignored the 13 CDFN sources for
which the redshift estimate is not based on two or more
emission/absorption lines (see Barger et al. 2003). No additional high
quality redshifts were obtained by cross-correlating the X-ray catalog
of Alexander et al. (\cite{alex03}) with the two recently published
spectroscopic catalogs of the ACS-GOODS survey in the CDFN (Cowie et
al. \cite{cowie04}, Wirth et al. \cite{wirth04}). The final considered
catalog includes 252 sources, corresponding to a spectroscopic
completeness of $\sim 50$\%.  We estimated the typical redshift errors
(not quoted in Barger et al. 2003) by comparing the common redshifts
with high quality in the catalogs of Barger et al. (\cite{barger03})
and Cohen et al. (\cite{cohen00}). We found that the measurements in
the two catalogs are in very good agreement, with essentially zero
offset and a dispersion of $\sigma(\Delta z)\lesssim 0.002$,
indicating that the redshift accuracy in each catalog is better than
this value. The redshift distribution for the considered spectroscopic
sample is shown in Fig.~\ref{zcdfn}. As in the case of the CDFS
redshift distribution, several redshift spikes can be immediately
identified, the most prominent of which at $z\sim 0.85$ and $z\sim
1.02$ (see Barger et al. 2003). 

\begin{figure}[t]
\includegraphics[width=9cm]{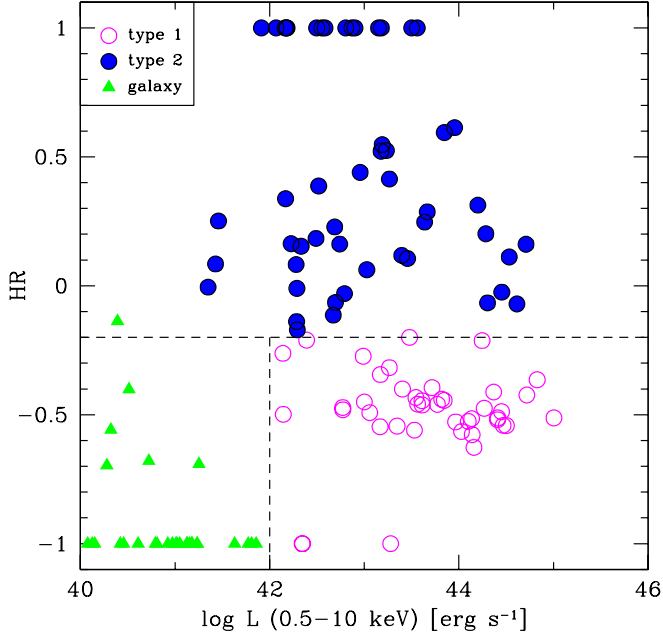}
\caption{The ``classification diagram'', i.e. hardness-ratio
vs. observed 0.5-10 keV luminosity, for the CDFS sources.}
\label{cls}
\end{figure}

Although the general shape of the
CDFN redshift distribution peaks at $z\sim 0.7-0.8$, similarly to that
observed in the CDFS (see e.g. the smoothed curves in Fig.~\ref{zcdfs}
and ~\ref{zcdfn}), a few differences can be noticed between the
two. One obvious effect is produced by the several spikes which trace
structures at different redshifts. More interestingly, the fraction of
low redshift sources is higher in the CDFN than in the CDFS. As an
example, 28\% of CDFN sources lay at $z<0.5$, while the corresponding
fraction in the CDFS is 17\%. This difference can be readily explained
by the deeper CDFN exposure, which is able to pick up the faint X-ray
emission of nearby normal and starburst galaxies (see the next Section
and the insets of Fig.~\ref{zcdfs} and ~\ref{zcdfn}). The X-ray flux
and R band magnitude distributions for the CDFN sources with good
redshift estimate considered in this paper are shown in the upper
panel of Fig.~\ref{ftdist} and Fig.~\ref{rdist}, respectively. Due to
the higher exposure time, in the CDFN the source flux distribution has
a larger fraction of objects at faint fluxes ($f_{0.5-10 keV}\lesssim
10^{-15}$ cgs) with respect to that observed in the CDFS \footnote{It
is worth noting that the 0.5-10 keV flux of the faintest CDFS sources
($f_{0.5-10 keV}\lesssim 10^{-16}$ cgs) is likely to be underestimated
in Fig.~\ref{ftdist}. Indeed, since in the CDFS no X-ray photometry
was performed in the total X-ray band, the 0.5-10 keV flux is obtained
by simply summing the flux in the soft and in the hard
band. Therefore, for sources detected in the soft band only (most of
which are at the faintest fluxes), the 0.5-10 keV flux simply
corresponds to the 0.5-2 keV flux, and some residual flux above 2 keV
is lost. On the contrary, in the CDFN the X-ray photometry has been
performed also in the total 0.5-10 keV band even for sources not
detected in the 2-10 keV band, whose total flux is then always higher
than the soft flux.}. As mentioned above, most of these faint sources
are classified as galaxies. The R-band magnitude distributions are
instead more similar, with most of the spectroscopically confirmed
sources in the range $19<R<24$ in both samples, confirming that the
spectroscopic observations have been equally deep in both fields.

\begin{figure}[t]
\includegraphics[width=9cm]{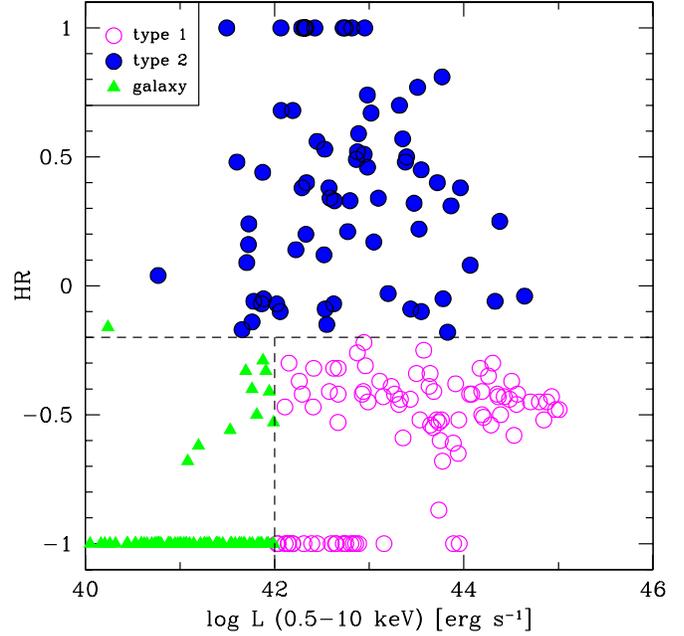}
\caption{Same as Fig.~\ref{cls} but for CDFN sources.}
\label{cln}
\end{figure}

\section {Source classification}

In order to measure the clustering properties of different
populations, we classified our sources following the scheme presented
by Szokoly et al. (\cite{szoko04}) for CDFS sources, where X-rays are
the main tool to infer informations on the physical nature of each
object. We somewhat simplified that scheme by avoiding the luminosity
distinction between type-2 AGN/QSOs and between type-1 AGN/QSOs. Our
adopted classification scheme can be then summarized as follows:

\begin{tabular}{lcl}
type-1 AGN &:& $HR<-0.2$ $and$ log$L_{0.5-10}\geq42$\\
type-2 AGN &:& $HR\geq-0.2$\\
galaxy    &:&  $HR<-0.2$ $and$ log$L_{0.5-10}<42$,\cr
\end{tabular}

\noindent
where $HR=(H-S)/(H+S)$ is the X-ray hardness ratio, i.e. the
difference between the hard ($H$) and soft ($S$) band counts
normalized to the total counts, and $L_{0.5-10}$ is the observed
0.5-10 keV luminosity in units of erg s$^{-1}$.

The cut at $HR=-0.2$ between type-1 and type-2 AGN is motivated by the
fact that most of the AGN with broad optical lines (31/32) lay below
this limit, while the majority of narrow line AGN (16/21) are found
above it. The adopted classification scheme is admittedly crude, but
it can be considered a reasonable approach when dealing with sources
with faint optical spectra, for which detailed line diagnostics is
difficult.

To keep a uniform classification criterion in the two field, we
applied the above scheme also to CDFN sources (see also Hasinger
2003).  As a consistency check, we computed the X-ray hardness ratio
for CDFN sources based on the soft and hard counts presented in the
Alexander et al. (\cite{alex03}) catalog, and verified that also for
this sample objects with broad optical lines have $HR\lesssim-0.2$ as
in the CDFS.

\begin{figure}[t]
\includegraphics[width=9cm]{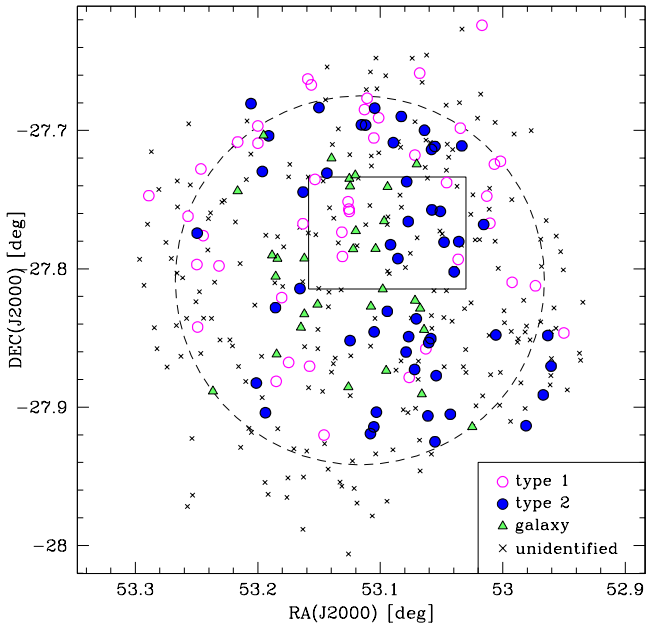}
\caption{Distribution on the sky of CDFS sources with robust redshift
measurements. Different source classes are represented
with different symbols as labeled. The box indicates the $6.7 \times
4.8$ arcmin region covered by the K20 survey (Cimatti et
al. \cite{cima02}). The dashed circle of 8 arcmin radius is the region
with higher ($\sim 50\%$) spectroscopic completeness.}
\label{imas}
\end{figure}

\begin{figure}[t]
\includegraphics[width=9cm]{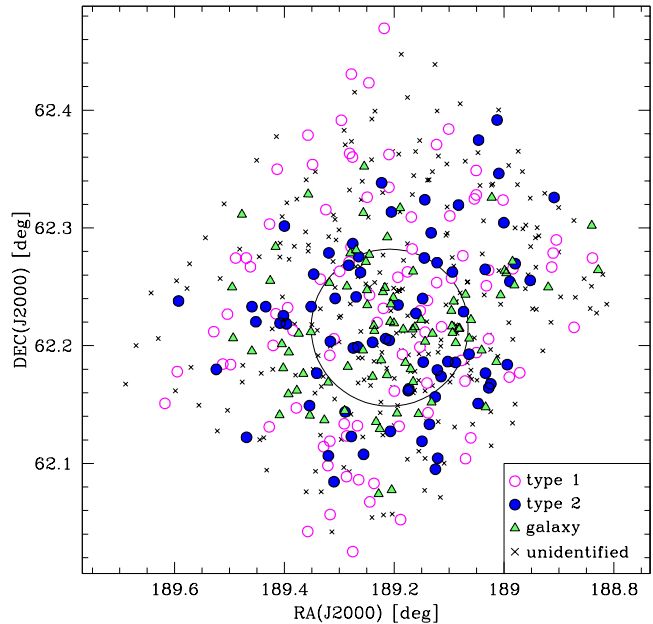}
\caption{Same as Fig.~\ref{imas} but for CDFN sources. Symbols are as
in the previous Figure. The 4 arcmin radius circle approximately shows
the area covered by the Hubble Deep and Flanking fields (Cohen et
al. \cite{cohen00}).}
\label{iman}
\end{figure}

The adopted ``classification diagram'', i.e. the hardness ratio vs.
X-ray luminosity plot, is shown in Fig.~\ref{cls} and \ref{cln} for
the CDFS and CDFN sources, respectively, and the classification
breakdown is shown in Table~1 (only sources with $L_{0.5-10}>10^{40}$
erg s$^{-1}$ are considered, see Section~4.2). We point out that the
significantly higher fraction of galaxies found in the 2Msec CDFN with
respect to the 1Msec CDFS is due to the twice longer exposure of the
CDFN. As shown in Fig.~\ref{cln}, the line at log$L_{0.5-10}=42$
appears to sharply divide a smooth source distribution into two
distinct classes (galaxies and type-1 AGN). It is therefore likely
that each class contain some misclassified objects. Indeed, part of
the soft sources with log$L_{0.5-10}<42$ might harbour a low
luminosity AGN and, on the other hand, galaxies with intense star
formation might have X-ray luminosities exceeding
log$L_{0.5-10}=42$. Nonetheless, the fraction of misclassified objects
should be of the order of a few percent in each class and therefore we
do not expect any significant impact on our clustering measurements.

\begin{table}
\caption{Source classification breakdown. Only sources with
$L_{0.5-10}>10^{40}$ erg~s$^{-1}$ are considered.}
\begin{tabular}{rccc}
\hline \hline
Sample& Type 1& Type 2& Gal\\
\hline
CDFS& 45& 52& 27\\
2Msec CDFN& 89& 71& 80\\
\hline
1Msec CDFN& 79& 60& 37\cr
\hline
\end{tabular}
\end{table}

The spatial distributions of the X-ray sources in the CDFS and CDFN as
a function of their spectroscopic classification are shown in
Fig~\ref{imas} and \ref{iman}, respectively. As it is evident in
Fig~\ref{iman}, most of the CDFN galaxies are found in the center of
the field, where the X-ray sensitivity is highest. When applying the
above classification scheme to the 189 CDFN sources with robust
redshift measurement detected in the first 1Msec exposure (Brandt et
al. \cite{brandt01}, Barger et al. \cite{barger02}), we found that,
while the number of AGN drops by $\sim 15\%$, the number of galaxies
drops by more than a factor of $\sim 2$, i.e. from 80 to 37. Then,
when accounting for the different spectroscopic completeness, the
number of galaxies found in the 1Msec CDFN is in agreement with that
found in the CDFS.  We also caution the reader that the ratio between
type-2 and type-1 AGN one might derive from Table~1 is a lower limit
rather than the real ratio in these deep X-ray fields: first of all,
the optical identifications are largely incomplete and the fraction of
type-2 AGN is expected to be higher among unidentified sources, which
are on average harder than those already identified (we indeed
verified that in both fields the type-2/type-1 ratio increases towards
faint R magnitudes); second, the number of obscured sources
misclassified as type-1, as defined on the basis of the here adopted
classification, is likely to be higher than the number of unobscured
sources misclassified as type-2. This can be seen for example in
Fig.~9 of Tozzi et al. (\cite{tozzi01}) where it is shown how the
observed hardness ratio decreases with redshift for a given value of
the obscuring column density $N_H$. In Gilli et al. (\cite{gilli03})
we classified an X-ray source as AGN with slightly different criteria
from those adopted here. In particular, we considered to be AGN those
sources satisfying at least one of the following conditions:
$L_{0.5-10}>10^{42}$ erg~s$^{-1}$, $HR>0$, $f_x/f_R>0.1$, where
$L_{0.5-10}$ is the observed 0.5-10 keV luminosity and $f_x/f_R$ is
the ratio between the 0.5-10 keV flux and the R band flux (see Section
4.1 of Gilli et al. 2003 for details). We verified that the two
classification criteria provide very similar results. Indeed, $\sim
97\%$ of the sources classified as AGN by one method are also
classified as AGN by the other.

\section {The spatial correlation function}

\subsection {Analysis techniques}

The most widely used statistics to measure the clustering properties
of a source population is the two point correlation function $\xi(r)$,
defined as the excess probability of finding a pair with one object in
the volume $dV_1$ and the other in the volume $dV_2$, separated by a
comoving distance $r$ (Peebles 1980):

\begin{equation}
dP = n^2[1+\xi(r)]dV_1dV_2
\end{equation}

A related quantity, which is what we actually measure in this paper,
is the so-called projected correlation function:

\begin{equation}
w(r_p) = \int_{-r_{v0}}^{r_{v0}} \xi(r_p, r_v)dr_v,
\end{equation}

where $\xi(r_p, r_v)$ is the two point correlation function expressed
in terms of the separations perpendicular ($r_p$) and parallel ($r_v$)
to the line of sight as defined in Davis \& Peebles (\cite{dp83}) and
applied to comoving coordinates. The advantage of using the integral
quantity $w(r_p)$ rather than directly estimating the two point
correlation function in redshift space $\xi(s)$ is that $w(r_p)$ is
not sensitive to distortions introduced on small scales by peculiar
velocities and errors on redshift measurements.

If the real space correlation function can be approximated by a
powerlaw of the form $\xi(r)=(r/r_0)^{-\gamma}$ and $r_{v0}= \infty$ then
the following relation holds (Peebles \cite{peeb80}):

\begin{equation}
w(r_p) =A(\gamma) r_0^{\gamma} r_p^{1-\gamma},
\end{equation}

where $A(\gamma)=\Gamma(1/2)\Gamma[(\gamma-1)/2]/\Gamma(\gamma/2)$
and $\Gamma(x)$ is the Euler's Gamma function. $A(\gamma)$
increases from 3.68 when $\gamma=1.8$ to 7.96 when $\gamma=1.3$.

\begin{figure}
\includegraphics[width=9cm]{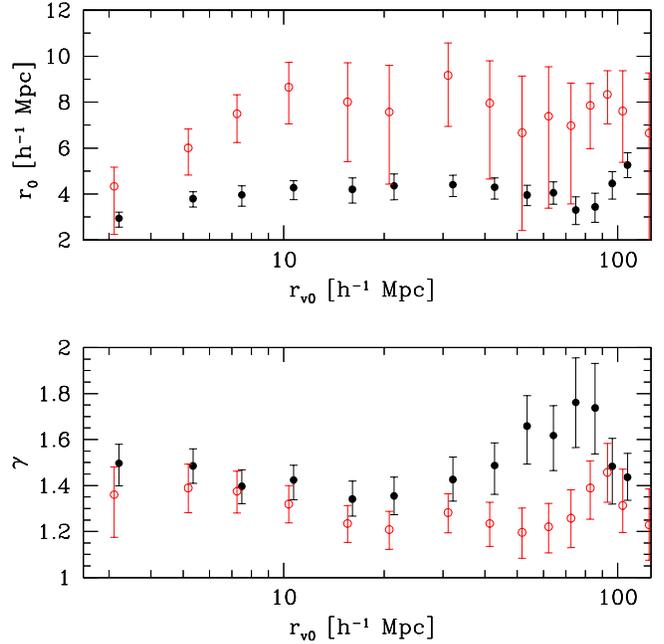}
\caption{Measured correlation length $r_0$ and slope $\gamma$ as a
function of $r_{v0}$, i.e. the integration limit on $w(r_p)$ (see
Eq.3), for the total samples in the CDFN (filled circles) and CDFS
(open circles). We choose $r_{v0}=10\:h^{-1}$ Mpc as our integration
radius. For lower $r_{v0}$ values the correlation signal is not fully
recovered, while for higher values the noise increases.}
\label{rov}
\end{figure}

A practical integration limit $r_{v0}$ has to be chosen in Eq.~2 in
order to maximize the correlation signal. Indeed, one should avoid too
large $r_{v0}$ values which would mainly add noise to the estimate of
$w(r_p)$. On the other hand too small scales, comparable with the
redshift uncertainties and with the pairwise velocity
dispersions, (i.e. the dispersion in the distribution of the relative
velocities of source pairs), should also be avoided since they would
not allow to recover the whole signal. A redshift uncertainty of
$\Delta z \lesssim 0.002$ (the typical value observed in our samples)
corresponds to comoving scales below $6.7\:h^{-1}$ Mpc at all
redshifts. The average velocity dispersion measured by Cohen et
al. (\cite{cohen00}) for the redshift spikes observed in the Hubble
Deep and Flanking fields is of the order of 400 km $s^{-1}$,
corresponding to $\Delta z\sim 0.002$ at $z\sim 0.7$. At these
redshifts the pairwise velocity dispersion should be of the same
order. Indeed, the value measured in the local Universe ($500-600$ km
$s^{-1}$; Marzke et al. \cite{marzke95}, Zehavi et al. \cite{zehavi02})
is expected to decrease by $\sim 15\%$ at a redshift of 0.7 (see
e.g. the $\Lambda$CDM simulations by Kauffmann et
al. \cite{kauff99}). We further checked that the velocity dispersion
measured for the redshift structures of X-ray sources in the CDFS and
CDFN corresponds typically to $\lesssim 10\:h^{-1}$ Mpc. To search for
the best integration radius $r_{v0}$ we measured $w(r_p)$ for the CDFS
and CDFN total samples for different $r_{v0}$ values ranging from 3 to
$100\:h^{-1}$ Mpc. The obtained correlation length and slope as a
function of $r_{v0}$ are shown in Fig.~\ref{rov}. We note that $r_{0}$
decreases for $r_{v0}$ values smaller than $10\:h^{-1}$ Mpc, showing
that the signal is not fully recovered. For $r_{v0}$ values greater
than $10\:h^{-1}$ Mpc $r_{0}$ does not vary significantly, but the
errorbars are higher. This behaviour, which is more evident for the
CDFS sample, is similar to that observed by Carlberg et
al. (\cite{carl00}) for the galaxies in the CNOC2 sample (Yee et
al. \cite{yee00}). The slope of the correlation is rather constant
over most of the $r_{v0}$ range. For the CDFN sample a steepening of
$\gamma$ is observed at $r_{v0}=50-90\:h^{-1}$ Mpc. However, at these
large radii, the errors are large and the measured slope is consistent
within $<2\sigma$ with the value obtained for $r_{v0}=10\:h^{-1}$
Mpc. We therefore consider the observed steepening as a fluctuation
which is not statistically significant and in the following we will
fix $r_{v0}$ to $10\:h^{-1}$ Mpc.


To measure $\xi(r_p, r_v)$ we created random samples of sources in 
our fields and measured the excess of pairs at separations $(r_p, r_v)$
with respect to the random distribution.
We used the minimum variance estimator proposed by Landy \& Szalay (1993),
which is found to have a nearly Poissonian variance:

\begin{equation}
\xi(r_p, r_v) = \frac{aDD(r_p, r_v)-2bDR(r_p, r_v)+RR(r_p, r_v)}{RR(r_p, r_v)},
\end{equation}

where DD, DR and RR are the number of data-data, data-random and
random-random pairs at separations $r_p \pm \Delta r_p$ and $r_v \pm
\Delta r_v$, $a=n_r(n_r-1)/n_d(n_d-1)$ and $b=(n_r-1)/2n_d$, where
$n_d$ and $n_r$ are the total number of sources in the data and random
sample, respectively.

Both the redshift and the coordinate $(\alpha, \delta)$ distributions
of the identified sources are potentially affected by observational
biases. In particular, the redshift distribution may be biased by the
presence of a limiting magnitude beyond which spectroscopic redshifts
can not be obtained. The $(\alpha, \delta)$ distribution, on the other
hand, is affected by at least two biases: the X-ray bias, due to the
non-uniform X-ray sensitivity limits over the field of view, and the
spectroscopic bias, due to the positioning of the masks within the
field and of the slits within the masks. For this reason special care
has to be taken in creating the sample of random sources. The
redshifts of these sources were randomly extracted from a smoothed
distribution of the observed one. This procedure should include in the
redshift selection function the same biases affecting the observed
distribution. We assumed a Gaussian smoothing length $\sigma_z = 0.3$
as a good compromise between too small smoothing scales (which suffer
from significant fluctuations due to the observed spikes) and too
large scales (where on the contrary the source density of the smoothed
distribution at a given redshift might be not a good estimate of the
average observed value). We verified that our results do not change
significantly when using a smoothing length in the range $\sigma_z =
0.2-0.4$. The smoothed redshift distributions adopted for our
simulations, shown in Fig.~\ref{zcdfs} and \ref{zcdfn} for the CDFS
and CDFN, respectively, have very similar shapes peaking at
$z\sim0.7$. We assumed that the clustering amplitude is constant with
redshift and did not try to estimate clustering variations at
different redshifts. Indeed, the clustering signal in a given redshift
interval will strongly depend on small variations in the choice of the
interval boundaries, which might include or exclude prominent redshift
spikes from the interval, hence producing extremely high fluctuations
in the $r_0$ vs. $z$ measurements. Since the X-ray sensitivity varies
across the field of view, in particular with off-axis angle, we
checked if there are significant differences in the redshift
distribution of sources as a function of their off-axis angles. In
particular we compared the distributions of sources inside and outside
a given off-axis angle with a Kolmogorov-Smirnov (hereafter KS)
test. We repeated the KS test for several source subsamples (e.g. AGN,
galaxies) in the CDFS and CDFN and for different off-axis angles. With
the exception of the galaxies in the CDFN, for which the average
redshift at off-axis angles below 4 arcmin is found to be
significantly higher than that outside this region, we do not find any
significant difference in the other subsamples. In the following we
will then generate the redshift distribution for the random samples by
simply smoothing the total distribution observed in each
subsample. The case of CDFN galaxies will be discussed in detail in
Section 4.2.2.
 
The coordinates ($\alpha,\delta$) of the random sources were extracted
from the coordinate ensemble of the real sample, thus reproducing on
the random sample the same uneven distribution on the plane of the sky
of the real sources (e.g. in both the CDFS and CDFN the X-ray sources
were identified preferentially at the center of the field). This
procedure, if anything, would dilute the correlation signal, since it
removes the effects of angular clustering. We note however that we do
not expect a strong signal from angular clustering in these deep
pencil-beam surveys, where the radial coordinate spans a much broader
distance than the transverse coordinate and the clustering signal
should be dominated by redshift clustering (see the tests with
random coordinates in the next section).

The source density adopted in the random samples is a factor of
50-100 larger than that of the data sample depending on its size.
More details on the chosen way to construct the random source sample,
as well as several checks on its validity will be discussed in the
next Section.

We binned the source pairs in interval of $\Delta{\rm log}\,r_p$=0.4 and
measured $w(r_p)$ in each bin. The resulting datapoints were then
fitted by a power law of the form given in Eq.~3, and the best fit
parameters $\gamma$ and $r_0$ were determined via $\chi^2$
minimization. Given the small number of pairs which fall into some
bins (especially at the smallest scales), we used the formulae of
Gehrels (\cite{gehre86}) to estimate the 84\% confidence upper and
lower limits, containing the 68\% confidence interval (i.e. $1\sigma$
errorbars in Gaussian statistics). It is well known that Poisson
errorbars underestimate the uncertainties on the correlation function
when source pairs are not independent, i.e. if the considered objects
generally appear in more than one pair. In the samples considered
here, this is indeed the case at scales $r_p\gtrsim 1\:h^{-1}$ Mpc. On
the other hand, bootstrap resampling techniques (e.g. Mo, Jing \&
B\"orner \cite{mo92}), which are often used to circumvent this
problem, may substantially overestimate the real uncertainties.
We tested bootstrap errors for our samples, finding that the
uncertainties on the correlation function parameters increase by a
factor of $\sim 2$ with respect to the Poissonian case. In the
following we will simply quote $r_0$ and $\gamma$ together with their
$1\sigma$ Poisson errors, bearing in mind that the most likely
uncertainty lay between the quoted number and its double.

\subsection {Results}

\subsubsection{CDFS}

We first considered the correlation function of all CDFS sources
regardless of their classification. We excluded from the sample only
stars and extended X-ray sources associated to galaxy
groups/clusters. In addition we excluded from our calculations 3 low
luminosity sources with $L_{0.5-10}<10^{40} $ erg~s$^{-1}$, in which
the X-ray emission might be due to a single off-nuclear Ultra Luminous
X-ray source in the host galaxy (ULX, see e.g. Fabbiano 1989) rather
than to the global star formation rate or to the active nucleus. We
note that Hornschemeier et al. (\cite{horn04}) found 10 ULX
candidates, all of them with $L_{0.5-10}\lesssim10^{40} $
erg~s$^{-1}$, in the combined CDFS + CDFN sample covered by the GOODS
survey. Although ULX likely do not represent the whole source
population below $10^{40} $ erg~s$^{-1}$, we nevertheless prefer to
apply this luminosity cut since only a few sources are lost and the
considered sample should be cleaner. Overall, we are left with a
sample of 124 sources.

\begin{figure}[t]
\includegraphics[width=9cm]{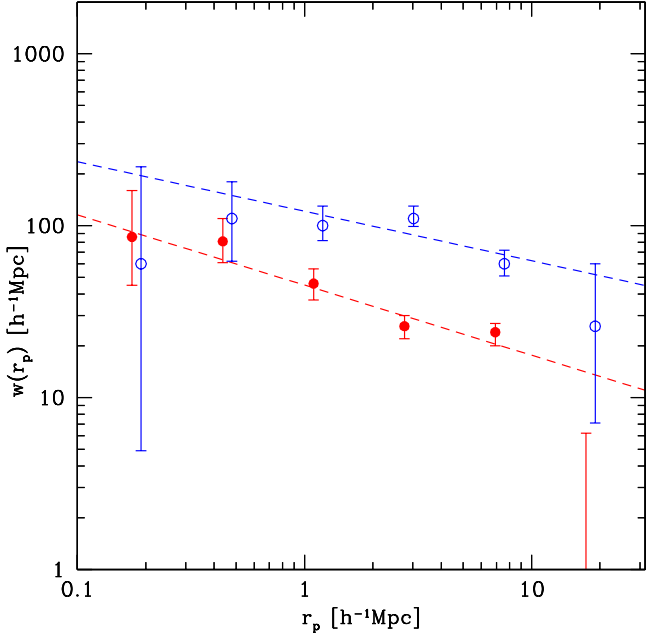}
\caption{Projected correlation functions for the total X-ray samples
in the CDFN (filled circles) and CDFS (open circles). Errors are
$1\sigma$ Poisson confidence intervals. The best fit power laws are
shown as dashed lines.}
\label{sall}
\end{figure}

The correlation function was measured in the redshift range $z=0-4$
(median redshift $\bar z\sim0.7$) and on scales $r_p=0.16-20\:h^{-1}$
Mpc. Here and in the following samples a power law fit is found
to be an adequate representation of the data. For the total CDFS
sample we obtained a fully acceptable value of
$\chi^2/dof=6.2/4$. The best fit correlation length is $r_0 = 8.6 \pm
1.2\: h^{-1}$ Mpc. The slope of the correlation, $\gamma=1.33 \pm
0.11$, is flatter than that commonly observed for optically selected
AGN and galaxies ($\gamma\sim1.6-1.8$, e.g. Le Fevre et al. 1996;
Croom et al. 2001). Based on the error on $r_0$ from this
two-parameters fit, we conservatively estimate the clustering signal
to be detected at the $\sim 7 \sigma$ level. We verified that
projected separations above $0.16\:h^{-1}$ Mpc correspond to angular
separations above 5 arcsec for sources in the considered redshift
range. Although the FWHM of the {\it Chandra} Point Spread Function
degrades with off-axis angle, it is still smaller than this value
within 8 arcmin from the center of the field, where $\sim 90\%$ of our
X-ray sources reside. Therefore, at the considered projected scales we
do not expect any strong bias against pairs with small angular
separations, which may artificially flatten the observed correlation
slope. In addition we checked if there is any bias against close pairs
because e.g. of the constraints on the slit positioning on the masks
used for optical spectroscopy. At any given separation we then
computed the ratio between the number of pairs in which both sources have
robust spectroscopic redshift and the total number of pairs at the
same angular separation. In fact, this ratio is rather constant,
decreasing by only $\sim 25\%$ at our smallest angular scales below
$\sim 20$ arcsec: this has some effects only at the smallest $r_p$
bins (at $z=0.7$, the median redshift of our sample, 20 arcsec
correspond to $\sim 0.17\:h^{-1}$ Mpc) where the clustering signal has
large uncertainties. Therefore no significant effects on the overall
best fit $\gamma$ value are expected. The projected correlation
function of the total CDFS sample is shown in Fig.~\ref{sall}.

\begin{figure}[t]
\includegraphics[width=9cm]{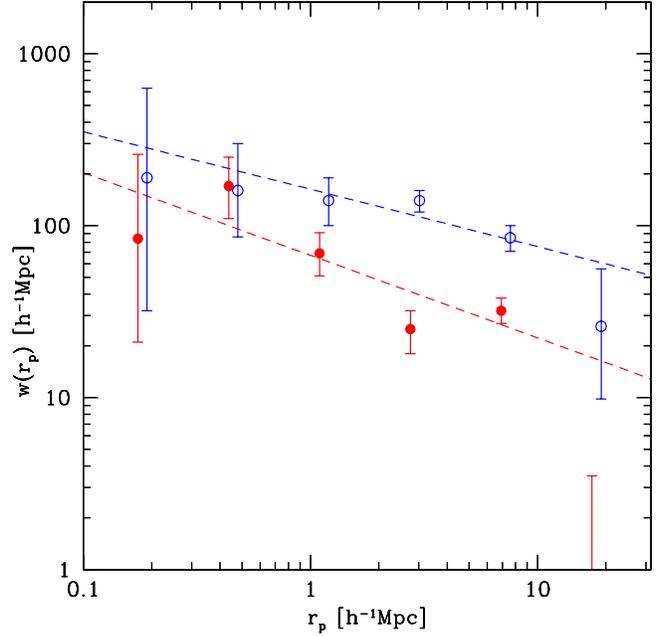}
\caption{AGN projected correlation functions in the CDFN (filled
circles) and CDFS (open circles). Errors are $1\sigma$ Poisson
confidence intervals. The best fit power laws are shown as dashed
lines.}
\label{sagn}
\end{figure}


We checked how much these results depend on the choice of the random
control sample. In particular we have relaxed the assumption of
placing the random sources at the coordinates of the real sources,
which might remove some signal due to angular clustering. As said
above it is not appropriate to randomly distribute the control sources
in the full field of view, since i) the X-ray sensitivity decreases
from the center to the outskirts of the field, and ii) the masks used
for optical spectroscopy have been placed preferentially in the center
of the field. As a first check we limited our analysis to the 110
sources within a circle with a radius of 8 arcmin from the center,
where the optical coverage is highest and the X-ray exposure map is
constant within $\sim 20\%$ across most of the field, with the
exception a few narrow stripes with lower sensitivity due to the gaps
among ACIS-I CCDs (see e.g. Fig.~3 of Giacconi et al. 2002).
Accordingly, the sources of the control sample were randomly placed
within this 8 arcmin circle. The best fit correlation length and slope
measured for this CDFS subsample were found to be
$r_0=9.0\pm1.1\:h^{-1}$ Mpc and $\gamma=1.38\pm0.14$, in excellent
agreement with the previously quoted values. We can therefore estimate
that the suppression in the clustering amplitude produced by the use
of the real coordinates is only of the order of a few percent.

As a further, more refined, check we created a probability
distribution map for the random sources, where the probability of
finding a source at a given position is proportional to the number of
real sources with measured redshift around that position. The map was
obtained by repeatedly smoothing the distribution of real sources on
the sky with a 20 arcsec boxcar (5 iterations). Random sources were
then placed in the field according to the created probability
map. This approach has the advantage of fully accounting for
observational biases, avoiding at the same time the removal of angular
clustering from the measured signal. Even in this case we found a
high correlation length and a flat slope ($r_0=9.1 \pm 1.0\:h^{-1}$
Mpc; $\gamma=1.36 \pm 0.10$), in agreement with the above derived
values. In the light of these checks, in the following we will
then simply place the random sources at the coordinates of the real
sources, considering for each AGN or galaxy subsample $only$ the
positions of the sources in that subsample.

Prompted by previous claims (Yang et al. \cite{yang03}), we checked if
there is any difference in the clustering properties of soft and hard
X-ray selected sources. The best fit parameters obtained for the 109
soft X-ray selected sources are $r_0=7.5\pm 1.4\:h^{-1}$ Mpc and
$\gamma=1.34 \pm0.14$, while for the 97 hard selected sources we
obtained $r_0=8.8 \pm 2.3\:h^{-1}$ Mpc and $\gamma=1.28 \pm
0.14$. Since the correlation length and slope are correlated, and
large uncertainties arise from the limited size of the samples, we
fixed $\gamma$ to a common value to best evaluate any possible
difference in the clustering amplitude. When fixing $\gamma$ to 1.3,
we found $r_0=7.5 \pm 0.6\:h^{-1}$ Mpc for the soft sample and
$r_0=9.1 \pm 0.8\:h^{-1}$ Mpc for the hard sample, which therefore
appears to be only marginally more clustered. \footnote{For
consistency with the other subsamples considered in this paper, we
quote in Table~2 the $r_0$ values obtained by fixing the slope to
$\gamma=1.4$ rather than to $\gamma=1.3$. Results are essentially
unchanged.}

\begin{figure}[t]
\includegraphics[width=9cm]{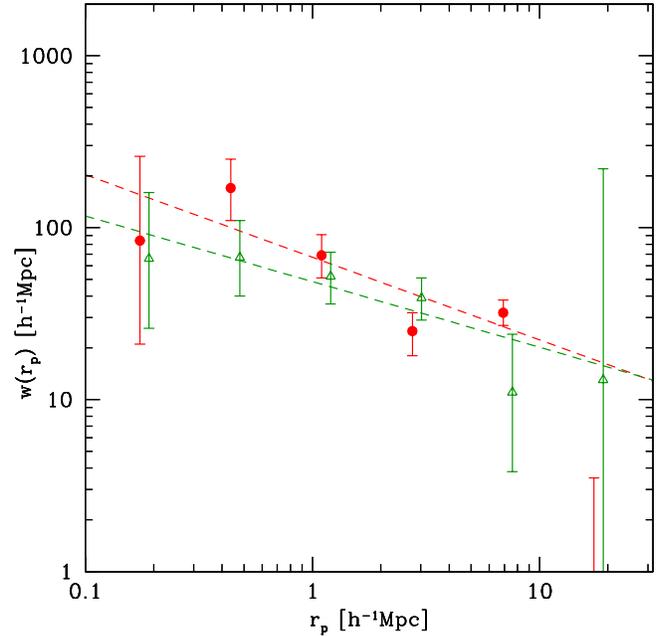}
\caption{Projected correlation functions for AGN (circles) and galaxies
(triangles) in the CDFN. Errors are $1\sigma$ Poisson confidence
intervals. The best fit power laws are shown as dashed lines.} 
\label{saga}
\end{figure}

\begin{figure}[t]
\includegraphics[width=9cm]{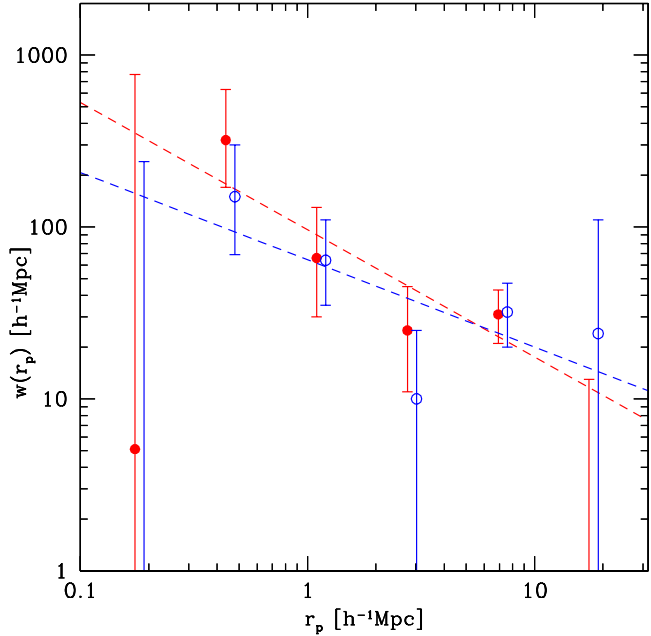}
\caption{Projected correlation functions for type 1 AGN (filled
circles) and type 2 AGN (open circles) in the CDFN. Errors are
$1\sigma$ Poisson confidence intervals. The best fit power laws are
shown as dashed lines.}
\label{satt}
\end{figure}

We then considered only the 97 sources classified as AGN finding best
fit values ($r_0=10.3 \pm 1.7\:h^{-1}$ Mpc, $\gamma=1.33 \pm 0.14$)
similar to those observed in the total sample (as it could be expected
since AGN represent the vast majority of the identified sources). The
AGN correlation function is shown in Fig.~\ref{sagn}. Furthermore, we
separated the total AGN sample into type 1 and type 2 AGN (45 and 52
objects, respectively) according to the classification diagram of
Section 3, without finding significant differences in their clustering
properties (see Table~2).
Because of the low statistics (only 27 objects) we cannot put
significant constraints to the galaxy correlation function.

Given the large errors introduced by low statistics, we fixed the
slope of the correlation function to $\gamma=1.4$ to search for any
difference in the $r_0$ values among different populations. The
adopted value is consistent with the average slopes measured in the
CDFS and in the CDFN. As expected, the $r_0$ values measured for the
various subsamples agree with those already obtained by assuming
$\gamma$ as a free parameter, but have smaller errors.
A summary of the measurements performed in this Section is given in
Table~2. We finally checked our results by fixing the slope of the
correlation to $\gamma=1.8$ which is the value commonly observed in
galaxy samples at low redshifts (Davis \& Peebles \cite{dp83},
Carlberg et al. \cite{carl00}): while the fit is significantly worse,
the best fit $r_0$ values increase by only 15\%.

\begin{table*}
\begin{center}
\begin{tabular}{lrccrcr}
\hline \hline
Sample& N& $\bar z$& ${\rm log}L_{0.5-10}$& $r_0$~~~~~~& $\gamma$& $r_0(\gamma=1.4)$\\
&&&&[$h^{-1}$ Mpc]&&[$h^{-1}$ Mpc]\\
\hline
\multicolumn{7}{c}{1Msec CDFS}\\
\hline
Total               &  124& 0.73&  43.0& $8.6 \pm 1.2$& $1.33 \pm 0.11$& $9.1 \pm 0.6$\\
Soft X-ray selected &  109& 0.73&  43.0& $7.5 \pm 1.4$& $1.34 \pm 0.14$& $7.6 \pm 0.7$\\
Hard X-ray selected &   97& 0.75&  43.3& $8.8 \pm 2.2$& $1.28 \pm 0.14$& $9.8 \pm 0.8$\\
AGN                 &   97& 0.84&  43.2& $10.3 \pm 1.7$& $1.33 \pm 0.14$& $10.4 \pm 0.8$\\ 
type 1              &   45& 1.03&  43.6& $9.1 \pm 3.3$& $1.46 \pm 0.33$& $10.1^{+1.8}_{-2.2}$\\    
type 2              &   52& 0.73&  42.8& $10.5 \pm 2.2$& $1.40 \pm 0.21$& $10.7^{+1.3}_{-1.6}$\\    
galaxies            &   27& 0.44&  41.0& \dots& \dots&\dots\\
\hline
\multicolumn{7}{c}{2Msec CDFN}\\
\hline
Total               &  240& 0.84&  42.4& $4.2 \pm 0.4$& $1.42 \pm 0.07$& $4.1 \pm 0.2$\\
Soft X-ray selected &  228& 0.84&  42.5& $4.0 \pm 0.4$& $1.42 \pm 0.08$& $4.1 \pm 0.3$\\
Hard X-ray selected &  149& 0.90&  43.0& $5.2 \pm 1.0$& $1.36 \pm 0.13$& $5.0 \pm 0.5$\\
AGN                 &  160& 0.96&  43.0& $5.5 \pm 0.6$& $1.50 \pm 0.12$& $5.1^{+0.4}_{-0.5}$\\ 
type 1              &   89& 1.02&  43.5& $6.5 \pm 0.8$& $1.89 \pm 0.23$& $5.6^{+0.8}_{-1.0}$\\    
type 2              &   71& 0.87&  42.7& $5.1 \pm 1.3$& $1.52 \pm 0.27$& $4.7^{+0.8}_{-1.0}$\\    
galaxies            &   80& 0.45&  41.3& $4.0 \pm 0.7$& $1.36 \pm 0.15$& $4.4^{+0.2}_{-0.6}$\\ 
\hline		 	      			   
\end{tabular}

\caption{Clustering measurements for different CDFS and CDFN
subsamples.  Errors are $1\sigma$ Poisson confidence levels. The
redshift range $z=0-4$ was considered for all the above samples except
for CDFN galaxies, where we used $z=0-1.5$. The considered sample,
number of objects in each sample and their median redshift and
luminosity are listed in Columns 1, 2, 3 and 4, respectively.  The
best fit correlation length and slope are quoted in Columns 5 and
6. The best fit correlation length obtained by fixing the slope to
$\gamma=1.4$ is quoted in Column 7.}
\end{center}
\end{table*}

\subsubsection{CDFN}

Most of the considerations made for the CDFS sample are also valid for
the CDFN sample. In particular a similar uneven distribution on the
field of the identified sources can be noticed in Fig.~\ref{iman}, so we
kept placing the sources of the random sample at the coordinates of
the real sources.

We first measured the correlation length for all the CDFN sources
excluding from our sample only objects with $L_{0.5-10}<10^{40} $
erg~s$^{-1}$ (i.e. possible ULX), leaving a final sample of 240
sources. Although no detailed information on the fraction of extended
sources is given in Alexander et al. (\cite{alex03}), the detection
procedure adopted for the 2Msec CDFN data should be optimized for 
point-like sources. In Alexander et al. (\cite{alex03}) it is indeed
mentioned that only a few sources are likely to be really extended;
their presence in the considered sample should therefore not affect
significantly our results.

We used again the redshift range $z=0-4$ since only two sources are
beyond $z=4$. The best fit parameters of the correlation function,
measured at a median redshift $\bar z\sim0.8$ are $r_0 = 4.2 \pm 0.4
\:h^{-1}$ Mpc and $\gamma=1.42 \pm 0.07$. Based on the error on
$r_0$, the clustering signal is then detected at the $\sim 10 \sigma$
level. While the slope is similar to that found in the CDFS, the
clustering amplitude is significantly smaller. The projected
correlation function of the total CDFN sample is shown in
Fig.~\ref{sall}, where it is also compared with that obtained for the
total CDFS sample.

Also in the CDFN we verified that the results do not change
significantly when limiting the calculation to the sources within 8
arcmin from the center (80\% of the full sample) and placing the
control sources randomly within this area.
Also in this field the clustering properties of various subsamples are
consistent with each other like for example those of soft and hard
X-ray selected sources (228 and 149 objects in the two subsamples,
respectively), and those of AGN (160 objects) and galaxies (80
objects). The best fit clustering parameters for the various samples
are quoted in Table~2. The projected correlation function of CDFN AGN
is compared with that of CDFS AGN in Fig.~\ref{sagn} and with that of
CDFN galaxies in Fig.~\ref{saga}.

As mentioned in Section 4.1, the average redshift of CDFN galaxies
seems to be higher in the center of the field than in the outer
regions. By means of a KS test we verified that the redshift
distributions of galaxies within and beyond 4 arcmin from the center
(38 and 42 objects, respectively) differ at $>3.5 \sigma$ level. To
check the possible effects on the measured correlation function, we
generated a first random sample by only considering the positions and
redshift distribution of the inner sources and a second random sample
by considering only the redshifts and coordinates of the outer
sources, and we finally pasted the two samples into one. In this way,
the outer sources of the random sample have on average lower redshifts
than the inner sources, as observed in the real sample. The galaxy
correlation function measured using this refined random sample is
found to be in excellent agreement with the previous measurement.

Finally, we searched for any possible difference in the clustering
properties of type 1 AGN (89 objects) and type 2 AGN (71
objects). Although type 1s seem to have a higher best fit correlation
length and a steeper slope than type 2s ($r_0 = 6.5 \pm 0.8 \:h^{-1}$
Mpc and $\gamma=1.89 \pm 0.23$ vs $r_0 = 5.1 \pm 1.3 \:h^{-1}$ Mpc and
$\gamma=1.52 \pm 0.27$), the two subsamples agree within the errors
(Fig.~\ref{satt}).

Again, we checked our results by fixing $\gamma$ to 1.4. Although the
$r_0$ have now smaller errors, we did not find any additional
difference in the clustering properties of the various source
populations. Finally, we checked our results by fixing the correlation
slope to $\gamma=1.8$ finding that the measured $r_0$ values increase
by $\sim 15\%$ as also seen in the CDFS. A summary of the measurements
performed in this Section is given in Table~2.

\section{Discussion} \label{discussion}

\subsection{The variance of the clustering amplitude}

The X-ray exposure in the CDFN is twice that in the CDFS. It is
therefore possible, in principle, that different populations with
different clustering properties are being sampled in the two fields at
the respective limiting fluxes. Indeed, as it can be easily seen in
Table~2, the median luminosity for the total source populations of the
CDFN is lower than that of the CDFS. This effect is primarily due to
the raise of the galaxy population at very faint X-ray fluxes (see
Fig.~\ref{ftdist} and Section~3). The median luminosities for the AGN
samples are nonetheless very similar in the CDFS and in the CDFN. We
performed a test by measuring the correlation function only for the
CDFN sources already detected in the first Msec catalog (Brandt et
al. \cite{brandt01}), which should guarantee an equal X-ray depth for
the CDFS and CDFN samples. For the sample of 189 1Msec CDFN sources
with robust spectroscopic redshift we found essentially the same
correlation length and slope found in the total 2Msec CDFN
sample. Therefore, the variance in $r_0$ between the CDFS and the CDFN
cannot be ascribed to the different depth of the X-ray
observations. We note that the redshift selection function obtained
for the 1Msec CDFN is almost identical to that obtained for the CDFS.

Also, as shown in Section~2, no systematic differences appear in the
follow-up programs of optical spectroscopy, with optically
faint source being equally observed in both fields. As assessed by a
KS test, the R magnitude distributions for the sources in our two
samples (i.e. those with robust redshift measurements,
Fig.~\ref{rdist}) are indistinguishable, although there is a marginal
hint that the fraction of sources with $R>24$ is slightly higher in
the CDFS than in the 2Msec CDFN ($14\pm4\%$ and $9\pm2\%$,
respectively). When considering the R magnitudes of the CDFN sources
in the 1Msec catalog, these are distributed as in the CDFS (again
checked with a KS test) and the fraction of faint ($R>24$) sources is
identical to that of the CDFS. Therefore, the variance in the
clustering amplitude cannot be explained by differences in the optical
spectroscopy depth. As a final -- perhaps redundant -- test, it has been
directly checked that the clustering amplitude in the two fields does
not vary when considering only sources with $R<24$.

In addition, we checked the R-K colors of our sources. In both fields
AGN are on average redder than galaxies. Indeed, AGN follow galaxy
color tracks (see Szokoly et al. 2004 and Barger et al. 2003) but lay
at higher redshifts than galaxies, where galaxy tracks are
redder. This can be understood by considering that, since the majority
of the AGN have low luminosities and are in many cases obscured, the
optical light is dominated by the contribution of the host
galaxy. When comparing the R-K color distribution of the sources in
the CDFS and in the CDFN we observed a very similar shape. This,
combined with the uncertainties in the R-K color determination, does
not allow us to remark any possible difference between the two fields.

\begin{figure}
\includegraphics[width=9cm]{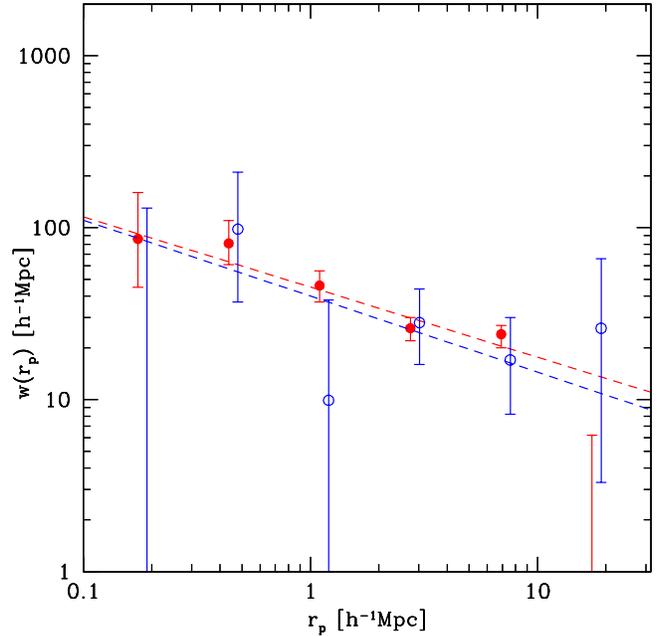}
\caption{Projected correlation function for the total CDFN sample (filled circles) 
and the CDFS sample obtained by excluding sources in the two spikes at z=0.67
and z=0.73 (open circles). Errors are $1\sigma$ Poisson confidence
intervals. The best fit power laws are shown as dashed lines.}
\label{nosp}
\end{figure}

We note that about 1/3 of the identified CDFS sources lay within the
two prominent spikes at $z=0.67$ and $z=0.73$. In the CDFN, although
several redshifts spikes are observed, there are no such prominent
structures. The two most populated spikes in the CDFN (at $z=0.84$ and
$z=1.02$) indeed contain only about 1/8 of the total identified
sources. As a check we measured the projected correlation function for
the total CDFS sample excluding the sources in the two redshift spikes
at $z=0.67$ and $z=0.73$, finding
$r_0=3.8^{+1.3}_{-2.7}\:h^{-1}$ Mpc and $\gamma=1.44\pm0.37$
($r_0=3.6\pm0.9\:h^{-1}$ Mpc when fixing $\gamma$ to 1.4) in good
agreement with the values measured for the total CDFN sample (see
Fig.~\ref{nosp}). We can therefore conclude that most of the
extra-clustering signal in the CDFS is due to these two structures. We
also verified that in the CDFN the clustering amplitude and slope do
not change significantly when removing the two most populated spikes
at $z=0.84$ and $z=1.02$.

We should also investigate if the observed variance might be induced
by the high spectroscopic incompleteness of the CDFN and CDFS samples.
When looking at the photometric redshifts (e.g. Zheng et al. 2004,
Barger et al. 2003), it can be easily shown that unidentified objects
lay on average at higher redshifts than spectroscopically identified
objects. The median redshift for the unidentified CDFS sources
(including photo-z and low quality spectro-z) is indeed 1.15 (1.40
when considering photo-z only), to be compared with 0.73, the median
redshift of the sources with high quality spectra. In the CDFN the
median redshift for unidentified sources is 1.17 (1.23 when
considering only photo-z), to be compared with the median value of
0.84 for the sources already identified. One of the most prominent
redshift spikes in the CDFN is at z=1.02 (see Fig.~\ref{zcdfn}), while
the most prominent structures in the CDFS are at $z\sim 0.7$. One
might then speculate that the CDFN spike at z=1.02 is more incomplete
than the CDFS spikes. Since at $z\sim0.7-1$ it is difficult to
identify sources with weak optical emission lines or sources with
absorption line dominated spectra, these should be the main population
missing from the spectroscopic samples. In the CDFS, where the
information on the optical spectra and classification is fully
available, we verified that the best fit parameters of the correlation
function do not vary significantly when excluding from the sample
sources with absorption line dominated spectra or only weak emission
lines. Therefore, spectroscopic incompleteness does not seem a viable
argument to explain the different clustering amplitude between the
CDFS and CDFN, which is rather due to genuine cosmic variance. We note
that large field to field variance might indicate a strong clustering
level, whose ``real'' amplitude can be assessed only with several
measurements on independent fields. In principle, the likelihood of
obtaining a given $r_0$ value for X-ray selected AGN in deep-pencil
beam surveys could be estimated by sampling several times a
cosmological volume obtained from N-body simulations, like e.g. the
``Hubble Volume Simulations'' by the Virgo Consortium (see Frenk et
al. 2000 and references therein). Unfortunately, this method requires
several assumptions on AGN formation and evolution within dark matter
halos and needs careful and extensive tests to evaluate all the
possible effects on the clustering amplitude of the considered
objects. Such an analysis is beyond the scope of this paper and will
be the subject of future work. 

An easier task, instead, is to see if the reported differences in
the number counts of the {\it Chandra} Deep Fields (e.g. Yang et
al. \cite{yang03}, Bauer et al. \cite{bauer04}) are consistent with
the fluctuations produced by the correlation lengths
$r_0=5-10\:h^{-1}$ Mpc that we measure. Very recently Bauer et
al. (\cite{bauer04}) have revisited the logN-logS relations in the
CDFS and CDFN finding general agreement between the two fields, the
maximum discrepancy (significant at the $\sim 4\sigma$ level) being
$\sim 40\%$ for hard sources at the faintest fluxes ($f_{2-10}\sim
4\:10^{-15}$ cgs; see their Fig.~5). Since we are considering sources
detected at the same limiting flux, the difference in the observed
surface density corresponds to a volume density difference of the same
entity. The expected cosmic variance in a given volume as a function
of the amplitude and slope of the correlation function can be
estimated using Eq.~3 of Somerville et al. (\cite{some04}), which is a
rearrangement of Eq.~60.3 by Peebles (\cite{peeb80}). Within comoving
effective volumes as those surveyed by each {\it Chandra} Deep Field
($\sim 2\:10^5 h^{-3}$ Mpc$^3$) and for a correlation slope
$\gamma=1.4$, the expected cosmic variance is 30\% and 50\% for
$r_0=5\:h^{-1}$ Mpc and $r_0=10\:h^{-1}$ Mpc, respectively.
Therefore, we conclude that the reported differences in the number
counts between the CDFS and the CDFN are fully consistent with the
correlation lengths measured in this paper.

\subsection{Comparison with clustering of other X-ray samples}

\begin{figure}
\includegraphics[width=9cm]{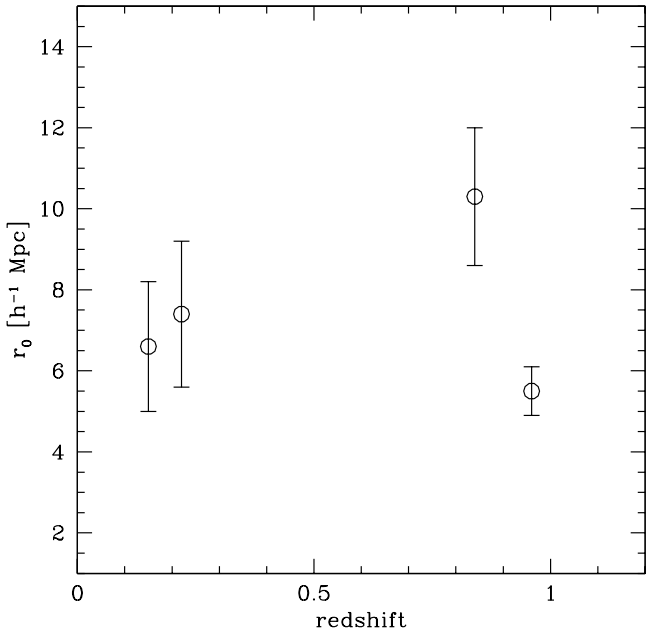}
\caption{Correlation length $r_0$ as a function of redshift for
different samples of X-ray selected AGN. From the lowest to the
highest redshift: RASS (Akylas et al. \cite{akylas00}); NEP (Mullis et
al. \cite{mullis04}); CDFS and CDFN (this work).}
\label{roz}
\end{figure}

Despite several efforts in the past years, only recently it has been
possible to directly measure the spatial clustering of X-ray selected
AGN. Carrera et al. (\cite{carrera98}) found only a $2\sigma$
detection in the ROSAT International X-ray Optical Survey (RIXOS,
Mason et al. 2000) on scales $<40-80\:h^{-1}$ Mpc. Interestingly, the
$2\sigma$ signal detected in the RIXOS refers to the subsample of
sources in the redshift range 0.5--1.0, where the biggest structures
in the CDFN and CDFS are also detected. The lack of clustering signal
at $z<0.5$ and $z>1$ might be due to the small volume sampled and to
the falling sensitivity of the RIXOS, respectively. More recently,
Mullis et al. (\cite{mullis04}) have measured the spatial correlation
function of soft X-ray selected AGN in the ROSAT NEP survey (their
clustering detection is at the $\sim 4\sigma$ level). Using the same
cosmology adopted here, they found a correlation length of
$r_0\sim7.4\pm1.8\:h^{-1}$ Mpc ($\gamma$ fixed to 1.8) for source
pairs at a median redshift $\bar z = 0.22$ and in the scale range
$5-60\:h^{-1}$ Mpc. Also, when accounting for the different cosmology
adopted here, the correlation length of the RASS sources at a median
redshift $z=0.15$, measured by Akylas et al. (\cite {akylas00}) through
angular clustering and Limber's equation, should be increased to
$r_0=6.6\pm 1.6\:h^{-1}$ Mpc
\footnote{At $z=0.15$ the average comoving separations in the
$\Lambda$ dominated cosmology adopted here are larger by $\sim 10\%$
with respect to the Einstein - De Sitter cosmology adopted by Akylas
et al. (2000).}. The correlation lengths measured at lower redshifts
in the NEP and RASS surveys are intermediate values between those
observed in the CDFS and in the CDFN. We stress that the comparison
between the {\it Chandra} Msec surveys and the NEP and RASS survey
should be done with the due care since they are sampling different
luminosity regimes, and AGN clustering is expected to be a function of
luminosity if this correlates with the mass of the dark halo in which
the AGN resides (e.g. Kauffmann \& Haenelt \cite{kauff02}). The median
0.5-10 keV luminosity of the NEP AGN (converted from the 0.5-2 keV
luminosity by assuming a spectrum with photon index 2) is indeed
log$L_{0.5-10}=44.4$, i.e. $\sim 20$ times higher than the median
luminosity in the CDFS and CDFN. The above consideration remarks how
the {\it Chandra} Msec surveys are sampling a population of AGN with
rather low luminosities, for which no information on clustering at
$z\sim 1$ was available so far. Another possible warning is that we
are comparing the soft X-ray selected AGN in the NEP and in the RASS
with the CDFS and CDFN AGN, which were selected both in the soft and
hard band. However we did not observe any significant difference in
the clustering properties of soft and hard X-ray selected AGN within
each field. In Fig.~\ref{roz} we show the correlation length of X-ray
selected AGN in the above mentioned surveys as a function of
redshift. Due to the variance in $r_0$ measured in the CDFS and CDFN,
no conclusion can be drawn on the evolution (if any) of the clustering
amplitude with redshift.

\subsection{Comparison with clustering of optically selected QSOs}

The best constraints on the clustering of optically selected QSOs have
been derived from the 2dF QSO Redshift Survey (2QZ, Croom et
al. 2001).  Based on a sample of $> 10^4$ objects Croom et
al. (\cite{croom01}) measured a QSO correlation length and slope of
$r_0=5.7 \pm 0.5 \:h^{-1}$ Mpc and $\gamma=1.56\pm 0.10$ at a median
redshift of $z=1.5$ and on scales $1-60 \: h^{-1}$ Mpc comoving (using
the same cosmology adopted here). In addition, thanks to the large
number of QSOs in their sample, Croom et al. (\cite{croom01}) were also
able to investigate the QSO clustering in different redshift
slices. The correlation length measured in their two lowest bins, at a
median redshift comparable with that of CDFS and CDFN AGN, is of the
order of $r_0=4.7 \pm 0.9 \:h^{-1}$ Mpc (for a fixed slope of
$\gamma=1.56$), which is comparable with the correlation length
measured for the CDFN AGN. Again, a fully meaningful comparison is
hampered by the different luminosity regimes sampled by the 2QZ and
the {\it Chandra} Msec surveys. Assuming a standard QSO SED (Elvis et
al. 1994), the characteristic absolute magnitude of 2QZ QSOs at
$z=0.9$, $M_{b_j}\sim-24.15$ (derived from the 2QZ luminosity function
of Croom et al. 2004), can be converted into an X-ray luminosity of
log$L_{0.5-10}=44.7$, well above the average values of CDFN and CDFS
AGN. In the local Universe the clustering of optical QSO has been
recently measured by Grazian et al. (\cite{grazi04}) by means of
the Asiago-ESO/RASS QSO survey (AERQS) which selects the most rare and
luminous objects with $B<15$ mag. These Authors measured a rather
high correlation length of $r_0=8.6 \pm 2.0 \:h^{-1}$ Mpc at a median
redshift of $\sim 0.1$ and on scales $1-30 \: h^{-1}$ Mpc comoving
(again for a fixed slope of $\gamma=1.56$). The average 0.5-10 keV
luminosity of their QSO sample can be estimated to be
log$L_{0.5-10}=44.4$. The AERQS and the 2QZ data have been compared
with QSO clustering evolution models (Matarrese et al. \cite{mata97},
Moscardini et al. \cite{mosc98}) based on the Press-Schechter
formalism for the evolution of the dark matter halo mass function. In
fact, Grazian et al. (\cite{grazi04}) and Croom et
al. (\cite{croom01}) derive a minimum mass for the dark matter halos
where QSO reside of $M_{DMH}\sim 10^{13}\:h^{-1}$ $M_\odot$. Due to
the present large uncertainties it is not yet possible to put
significant constraints to clustering evolution models with X-ray
selected AGN. We just note here that clustering of X-ray AGN is
consistent with models with $M_{DMH}\sim 10^{13}\:h^{-1}$ $M_\odot$ if
the low $r_0$ value measured in the CDFN is typical at $z\sim1$. On
the other hand, if the $r_0$ value measured in the CDFS has to be
considered as typical, then $M_{DMH}$ can be as high as
$10^{14}\:h^{-1}$ $M_\odot$.


\subsection{Comparison with galaxy clustering}

Gilli et al. (\cite{gilli03}) found that about $70-80\%$ of the high
significance peaks seen in the redshift distribution of K-band
selected sources in a sub-area of the CDFS (the area covered by the
K20 survey, see Fig.~\ref{imas}; Cimatti et al. 2002), have a
corresponding peak in the X-rays. This implies that X-ray and
K-band selected sources are tracing the same underlying
structures. Also, it might be speculated from these samples that AGN
clustering is similar to that of early type galaxies, whose detection
rate is higher in K-band rather than in optically selected
samples. The measurements of the spatial correlation function for the
AGN in the CDFS seem to be in agreement with this idea, since the
measured AGN correlation length is found to be similar to that of
Extremely Red Objects with $R-K>5$ (EROs) at $z\sim1 $, which are
thought to be the progenitors of early type galaxies (Daddi et
al. \cite{daddi01}). Such a high clustering amplitude is however not
observed for the AGN in the 2Msec CDFN, for which $r_0$ is of the
order of $5-6\:h^{-1}$ Mpc. If we then consider the AGN correlation
length to be in the range 5-10 Mpc, this is still consistent with AGN
at $z\sim1$ to be generally hosted by early type galaxies. Indeed Coil
et al. (2003) have recently measured the correlation length of a
sample of $\sim 2000$ R-band selected galaxies at $z=0.7-1.25$ in the
DEEP2 survey. With these good statistics they were able to obtain an
accurate measure of the correlation function of early-type and
late-type galaxies separately (the latter being more numerous by a
factor of $\sim 4$), finding $r_0=6.61\pm1.12\:h^{-1}$ Mpc for early
type galaxies and $r_0=3.17\pm0.54\:h^{-1}$ Mpc for late type
galaxies. Interestingly enough, on scales of $r_p=0.25-8\:h^{-1}$ Mpc,
i.e. very similar to those adopted in this paper, the slope of the
correlation function for early-type galaxies is found to be rather
flat, $\gamma=1.48\pm0.06$, in agreement with that measured for the
AGN in the CDFS and in the CDFN (note however that Guzzo et
al. (\cite{guzzo07}) found $\gamma=2.0\pm0.1$ for local early type
galaxies). On the contrary the correlation slope for late-type
galaxies is found to be significantly steeper
($\gamma=1.68\pm0.07$). To summarize, our results are consistent
with the idea that at $z\sim 1$ the population of AGN with typical
X-ray luminosity of $10^{43}$ erg~s$^{-1}$ is preferentially hosted by
early-type galaxies. However, other deep X-ray pointings in separate
fields are needed to measure the average clustering of X-ray
selected AGN and get more stringent results.


\section{Conclusions and future work} \label{conclusions}

We have measured the projected correlation function $w(r_p)$
of X-ray selected AGN and galaxies in the 2Msec {\it Chandra} Deep
Field North and in the 1Msec {\it Chandra} Deep Field South on scales
$\sim 0.2-10 \:h^{-1}$ Mpc. A significantly different amplitude for
AGN clustering has been observed in these $\sim 0.1$ deg$^2$ fields,
the correlation length $r_0$ measured in the CDFS being a factor of
$\sim 2$ higher than in the CDFN. The observed difference does not
seem to be produced by any observational bias, and is therefore likely
due to cosmic variance. In both fields the slope of the correlation
function is found to be flat ($\gamma\sim 1.3-1.5$), but consistent
within the errors with that measured for optically selected QSO (Croom
et al. \cite{croom01}). The extra correlation signal present in the
CDFS is primarily due to the two prominent spikes at $z=0.67$ and
$z=0.73$ containing about 1/3 of the identified sources. Indeed,
although significant redshifts spikes are also observed in the CDFN,
they are less prominent than those observed in the CDFS. In the CDFN
we were also able to measure the clustering properties of X-ray
selected galaxies, which have been found to be similar to those of AGN
in the same field.
Finally, within each field, we did not find significant differences
between the clustering properties of hard X-ray selected and soft
X-ray selected sources, or, similarly, between type-1 and type-2 AGN.

Significant improvements in the measurements of the AGN spatial
correlation function and then in the understanding of the large scale
structures in the X-ray sky is expected from the on going observations
of the Extended {\it Chandra} Deep Field South (E-CDFS, PI N. Brandt) and
of the COSMOS-XMM field (PI G. Hasinger). The E-CDFS is a deep-and-wide
survey consisting of 4 {\it Chandra} 250 ksec ACIS-I pointings
arranged in a square centered on the Msec CDFS. The final covered area
will be $\sim 0.3$ deg$^2$, i.e. a factor of 3 higher than that
covered by the Msec CDFS, with average sensitivities of $1\;10^{-16}$ erg
cm$^{-2}$ s$^{-1}$ in the soft band and $1\;10^{-15}$ erg cm$^{-2}$
s$^{-1}$ in the hard band. This will allow to significantly enlarge
the sample and reduce statistical uncertainties introduced by the
small CDFS field of view in the measurements of the clustering of
Seyfert-like AGN with average log$L_{0.5-10}=43$ erg s$^{-1}$. A
detailed study of clustering of high-luminosity X-ray selected AGN
will be instead performed by the wide area COSMOS-XMM survey,
consisting of a mosaic of 25 XMM short pointings (32 ksec each)
covering a total 2.2 deg$^2$ field with a sensitivity of $1\;10^{-15}$
erg cm$^{-2}$ s$^{-1}$ in the soft band and $4\;10^{-15}$ erg
cm$^{-2}$ s$^{-1}$ in the hard band. The two projects are
complementary and should constrain the clustering properties of X-ray
selected AGN as a function of redshift and luminosity.


\acknowledgements

We warmly thank Chris Mullis for useful discussions and for sharing with us
his results in advance of publication. The anonymous referee
is acknowledged for providing several comments which improved the
presentation of this work. RG acknowledges support from the Italian
Space Agency (ASI) under grant I/R/057/02.


\end{document}